# High-sensitivity sensing technology of 10 MHz radio frequency electric field based on Rydberg atoms*

HAN Yulong[1, 2, 3], SHAN Zhengye[1], ZHANG Kan[1], SUN Jinfang[1], ZHANG Lihua[2, 3, *], LIU Bang[2, 3, *], DING Dongsheng[2, 3]

1. School of General Education and Foreign Languages, Anhui Institute of Information Technology, Wuhu 241100, China
2. Laboratory of Quantum Information, University of Science and Technology of China, Hefei 230026, China
3. Anhui Province Key Laboratory of Quantum Network, University of Science and Technology of China, Hefei 230026, China

**Abstract**

To address the pressing need for high-sensitivity, miniaturized electric field sensors in the short-wave communication band (3 – 30 MHz), this study proposes and demonstrates a radio frequency electric field quantum sensing scheme based on cesium atomic Rydberg states and the electromagnetically induced transparency (EIT) effect. The scheme employs an all-infrared optical design, utilizing 852 nm probe light, 1470 nm dressing light, and 780 nm coupling light to efficiently prepare the $49P_{3/2}$ Rydberg state in a room-temperature cesium vapor cell, which is then probed non-destructively via ladder-type four-level EIT spectroscopy.

The 10 MHz RF field is non-resonant with adjacent Rydberg atomic energy levels, this work innovatively utilizes the alternating current (AC) Stark shift effect as the sensing mechanism. Real-time monitoring of the spectral shift of the Rydberg-EIT peak with the applied electric field strength enables direct electric field metrology. The evolution of the EIT spectrum was experimentally observed under electric fields ranging from 0 to 500 V/m: at weak fields (< 20 V/m), the spectrum primarily exhibits shifting and broadening; at moderate fields (20 – 100 V/m), oscillations indicative of dissipative time crystals emerge due to multi-Rydberg-level competition; at stronger fields, spectral splitting and modulation sidebands appear.

---

To further enhance the detection sensitivity for weak signals, a non-resonant superheterodyne detection technique was introduced. A local oscillator (LO) electric field at 10.1 MHz and the signal field at 10 MHz are simultaneously coupled into the atomic system. The atomic medium's amplified response to their beat note signal ($\Delta\omega$ = 100 kHz) down-converts the signal of interest to a lower, more easily detectable frequency domain. Precise measurement of the beat note signal strength enables high-sensitivity demodulation of the signal electric field.

Experimental results demonstrate outstanding sensing performance at 10 MHz: an electric field sensitivity of 31.0 $\mu V \cdot cm^{-1} \cdot Hz^{-1/2}$, a dynamic range of 65 dB, and an instantaneous bandwidth of approximately 0.6 MHz. This study develops a precise electric field measurement method for the MHz frequency band based on Rydberg atoms. Its centimeter-scale sensing unit overcomes the trade-off between size and sensitivity inherent in traditional antennas, offering a new technological pathway for miniaturized, high-sensitivity electric field sensors in applications such as short-wave communication, marine, and aeronautical navigation.

# 1 Introduction

Rydberg atoms feature outermost electrons in highly excited states. Their large polarizability and electric dipole moments[1] have attracted significant attention. In recent years, electrical measurement techniques based on Rydberg atoms have become a research hotspot[2-6]. Early Rydberg atom systems relied on field ionization detection[7]. However, this method destroys the Rydberg state during measurement, preventing continuous monitoring. In 2012, Sedlacek et al.[8] utilized the electromagnetically induced transparency (EIT) effect[9] and Autler-Townes (AT) splitting[10] in Rydberg atoms. They achieved non-destructive precision measurement of microwave electric fields in a thermal atomic vapor cell. This work laid the foundation for Rydberg atom quantum metrology. Since then, researchers have continuously worked to enhance sensor performance through novel architectures. In 2020, Jing et al.[3] proposed an atomic superheterodyne receiver scheme. This approach achieved a microwave electric field intensity measurement sensitivity of 55 $nV \cdot cm^{-1} \cdot Hz^{-1/2}$. Recently, Yan et al.[11] introduced a method based on Sagnac interference enhancement. Their technique reached a microwave field strength measurement sensitivity of 10.7 nV·cm$^{-1}$·Hz$^{-1/2}$. These advances fully demonstrate the unique advantages of Rydberg atom systems in electric field measurement. Their response to electromagnetic waves breaks the physical size limitations of traditional metal antennas. Furthermore, the response frequency can be continuously tuned from hundreds of megahertz to terahertz using lasers. Given these characteristics, this technology has found applications in electrical equipment discharge detection[12,13], broadband signal detection[14,15], digital communication[16], and sub-wavelength imaging[17,18].

Although Rydberg atoms enable broadband electric field measurement, early studies primarily focused on high-sensitivity measurements in the 300 MHz-300 GHz microwave band[3,11,19-21]. However, significant progress has recently been made in high-sensitivity measurements of low-frequency signals[22,23]. For instance, Liu et al.[4] achieved precision detection of 30 MHz low-frequency radio frequency (RF) electric fields using a non-resonant heterodyne method based on Rydberg atoms. The shortwave band (3-30 MHz) plays an irreplaceable role in long-distance communication, international point-to-point transmission, and radio, maritime, and aviation communication due to its ionospheric reflection properties[24-26]. However, traditional low-frequency electric field detection techniques rely on meter-scale antennas to achieve sufficient sensitivity. This requirement largely limits their practical application. Therefore, developing centimeter-scale compact megahertz electric field sensors with high sensitivity based on Rydberg atoms holds significant application value.

This study implements high-sensitivity measurement of 10 MHz RF electric fields using highly excited Rydberg states of cesium atoms based on EIT optical detection. Since RF electric fields in this frequency band cannot directly couple energy level transitions between Rydberg states, we employ the alternating current (AC) Stark shift effect as the detection mechanism. We achieve real-time recording of the electric field response by monitoring changes in the EIT spectrum relative to the applied electric field strength. To further enhance the detection sensitivity of the system, we introduce non-resonant heterodyne detection technology[3,4,27]. This technique amplifies weak signal responses using a local oscillator electric field. Experimental results indicate that the system achieves an electric field sensitivity of 31.0 μV·cm$^{-1}$·Hz$^{-1/2}$ for 10 MHz electric field measurements. Additionally, it exhibits a dynamic range greater than 65 dB.

# 2 Fundamental Principles

## 2.1 Experimental Setup

The experiment employs a ladder-type four-level Rydberg-EIT scheme. In this energy level system, $|1\rangle$ represents the ground state of the cesium atom, $|2\rangle$ and $|3\rangle$ denote two low-lying excited states, and $|4\rangle$ corresponds to the Rydberg state, as shown in Figure 1(a). The probe light and reference light, both with a wavelength of 852 nm, enter via parallel optical paths. They pass through a cylindrical quartz glass cesium vapor cell with a length of 70 mm and a diameter of 30 mm. The dressing light (1470 nm) and coupling light (780 nm) are spatially overlapped. They then pass through the cesium vapor cell in the direction opposite to the probe light. These three laser beams collectively excite the Rydberg atoms, establishing the atomic electric field detection region. The probe beam waist radius ($1/e^2$ radius) is approximately 200 μm, and the probe light power is 64 μW. The Rabi frequency is $\Omega_p$. This beam focuses within the cesium vapor cell to drive the transition of cesium atoms from the ground state $|6S_{1/2}\rangle$ to the excited state $|6P_{3/2}\rangle$. The dressing light, with a Rabi frequency of $\Omega_d$, enables the coupled transition between the two intermediate states $|6P_{3/2}\rangle$ and $|7S_{1/2}\rangle$. The dressing beam waist radius is approximately 500 μm, and the dressing light power is 16.8 mW. The coupling beam waist is approximately 500 μm, and the coupling light power is 1.5 W. The Rabi frequency is $\Omega_c \approx 2\pi\times4$ MHz. This beam primarily drives the transition from the excited state $|7S_{1/2}\rangle$ to the Rydberg state $|49P_{3/2}\rangle$.

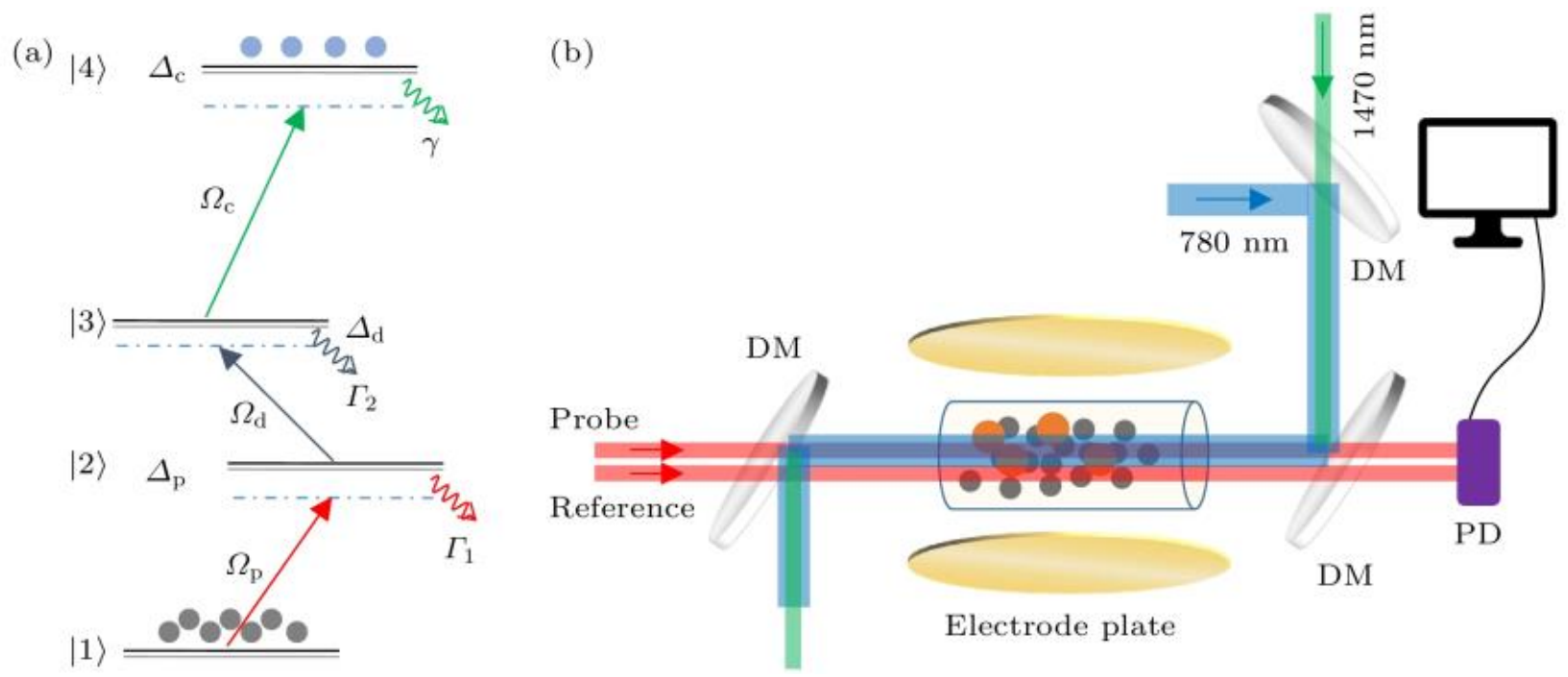

**Fig. 1 (a) Schematic diagram of a ladder-type four-level system comprising the ground state $|1\rangle$ of Cs atoms, two low-lying excited states $|2\rangle$ and $|3\rangle$, and the Rydberg state $|4\rangle$; (b) overview of the experimental setup. Two copper electrode plates are positioned parallel to each other on opposite sides of the cesium vapor cell with a separation of 4 cm. DM denotes a dichroic mirror, and PD represents a photodiode detector.**

Preparing cesium atom Rydberg states using a four-level structure avoids short-wave lasers with shorter wavelengths. This approach effectively suppresses ionization shielding caused by the photoionization of Cs atoms in the vapor cell[28,29]. The lasers used in the experiment are standard commercial models. They are readily available and do not require expensive and complex frequency-doubling laser systems. To prevent excessive charge accumulation at the edges of the plates, the experiment employed circular copper plates with a thickness of 3 mm and a diameter of 120 mm as electrodes.

These two electrodes were placed parallel to each other on both sides of the cesium cell. The distance between them was adjusted to 40 mm, and the cesium cell was fixed at the center of the plates. Two radio frequency (RF) signal sources used in the experiment provided the local oscillator (LO) signal and the signal under test, respectively. Due to a slight frequency detuning between the two signals, the generated electric field was coupled to the transmission line via an RF power splitter. It was finally applied to the electrode plates near the atomic vapor cell. Meanwhile, a balanced photodetector performed differential detection of the transmission difference between the probe and reference beams. This process extracted subtle changes in the EIT spectrum. An oscilloscope and a spectrum analyzer, controlled by a computer, completed the acquisition and recording of the photodetector signals.

## 2.2 Theoretical Framework

The system was modeled based on the ladder-type four-level diagram shown[30] in Figure 1(a). The detunings of the probe, dressing, and coupling lasers are denoted as $\Delta_p$, $\Delta_d$, and $\Delta_c$, respectively. The decay rates of atomic states $|2\rangle$, $|3\rangle$, and $|4\rangle$ are denoted as $\Gamma_1$, $\Gamma_2$, and $\gamma$, respectively. Using the rotating wave approximation, the Hamiltonian of the four-level system for light-atom interaction can be expressed as

$$H_I = \begin{aligned} &\hbar\Delta_p\sigma_{22} + \hbar(\Delta_p + \Delta_d)\sigma_{33} \\ &+\hbar(\Delta_p + \Delta_d + \Delta_c)\sigma_{44} \\ &-\hbar/2(\Omega_p\sigma_{12} + \Omega_d\sigma_{23} + \Omega_c\sigma_{34} + \text{H.c.}), \end{aligned} \tag{1}$$

Here, $\sigma_{ij} = |i\rangle\langle j|$ $(i, j = 1, 2, 3, 4)$ denotes the atomic transition operator. Considering spontaneous emission, the Lindblad equation governing the system can be expressed as

$$\dot{\rho} = -\frac{i}{\hbar}[H, \boldsymbol{\rho}] + \mathcal{L}(\Gamma_{21}) + \mathcal{L}(\Gamma_{32}) + \mathcal{L}(\Gamma_{43}), \tag{2}$$

In the equation, $\boldsymbol{\rho}$ denotes the density matrix of the system. The Lindblad operator is expressed as $L(\Gamma_{ij}) = \Gamma_{ij}/2\,(2\sigma_{ij}\rho\sigma_{ij} - \sigma_{ii}\rho - \rho\sigma_{ii})$, where $\Gamma_{21} = \Gamma_1$, $\Gamma_{32} = \Gamma_2$, and $\Gamma_{43} = \gamma$. To account for the Doppler effect induced by atomic thermal motion, we apply corresponding corrections to the detunings. These are defined as $\Delta_p = \Delta_p - k_p v$, $\Delta_d = \Delta_d + k_p v$, and $\Delta_c = \Delta_c + k_c v$. Here, $k_p$, $k_d$, and $k_c$ represent the corresponding wave vectors, while $v$ denotes the atomic velocity. We obtain the steady-state solution for the system density matrix by solving Equation (2). Subsequently, we derive the system susceptibility as follows:

$$\chi_{21} = -\int \frac{2N(v)|d_{21}|^2}{\hbar\varepsilon_0\Omega_p}\rho_{21}(v)\mathrm{d}v, \tag{3}$$

Here, $N(v) = N_0/u\sqrt{\pi}\exp(-v^2/u^2)$, where $u = (2k_B T/M)^{-1/2}$ represents the most probable speed, $N_0$ denotes the atomic density, and $d_{21}$ is the transition dipole matrix element between $|2\rangle$ and $|1\rangle$. Additionally, $k_B$ is the Boltzmann constant, $T$ is the temperature inside the cesium cell, and $M$ is the mass of a cesium atom. The absorption coefficient $\alpha$ is derived from the imaginary part of the susceptibility, expressed mathematically as $k_p\mathrm{Im}(\chi_{21})$. The transmittance is calculated using the formula $\exp(-\alpha l)$, where $l$ represents the length of the cesium cell.

# 3 Experimental Results and Analysis

As shown in Figure 1(b), the diameter of the copper plates exceeds that of the cesium cell. Neglecting edge effects, we approximate the internal environment of the cesium cell as a uniform electric field when voltage is applied. By experimentally observing the coupling light EIT signal, we detect the Stark level shift $\Delta$, thereby enabling electric field measurement based on Rydberg atoms. The Stark level shift $\Delta$ is an even function of the electric field $E$. Consequently, the direction of the Stark shift is independent of the polarity of the applied voltage. We use the ARC package[31] to calculate the dipole polarizability of atomic states and obtain the DC polarizability through fitting. Calculations indicate that the DC polarizabilities[32] of the cesium Rydberg states $|49P_{3/2}\rangle$ and $|49P_{1/2}\rangle$ exceed that of the $|7S_{1/2}\rangle$ state by more than three orders of magnitude. Therefore, in the experimental observations, we attribute all contributions to the Stark shift of the Rydberg states, ignoring the influence of the $|7S_{1/2}\rangle$ state polarizability. During the experiment, we fix the frequencies of the probe and coupling lights to resonate with their respective transitions. We obtain the EIT spectrum of cesium atoms by scanning the detuning $\Delta_c$ of the coupling light. Under a sinusoidal RF electric field at 10 MHz, we measure the variation of the EIT spectrum with electric field strength. The measurements cover an external field intensity range of 0-500 V/m with a step size of 1.25 V/m. As illustrated in Figure 2(a), when the RF electric field strength $E$ = 0 V/m, the system remains in a state without an external field. The measured Rydberg state EIT spectrum under these conditions serves as the reference spectrum. The primary and secondary peaks in this spectrum correspond to the $|49P_{3/2}\rangle$ and $|49P_{1/2}\rangle$ energy levels, respectively. Figure 2(b) displays the detailed spectral features of the $|49P_{3/2}\rangle$ state EIT spectrum under RF electric fields of varying intensities. Within the electric field strength range of 0-20 V/m, the EIT lines exhibit only Stark shifts and some broadening. This phenomenon is attributable to the collisional ionization effect of Rydberg atoms[33]. In the range of 20-100 V/m, the EIT lines show significant oscillations. Multiple experimental groups have reported that these oscillations result from dissipative time crystals caused by competition among multiple Rydberg levels[34-37]. As the field strength increases further, the main EIT peak gradually disappears, and modulation sidebands emerge. Low-frequency electric fields induce the formation of an electrostatic shielding layer on the inner wall of the cesium cell[38]. Establishing a stable charge distribution requires time, and the electric field shielding effect weakens as the frequency increases.

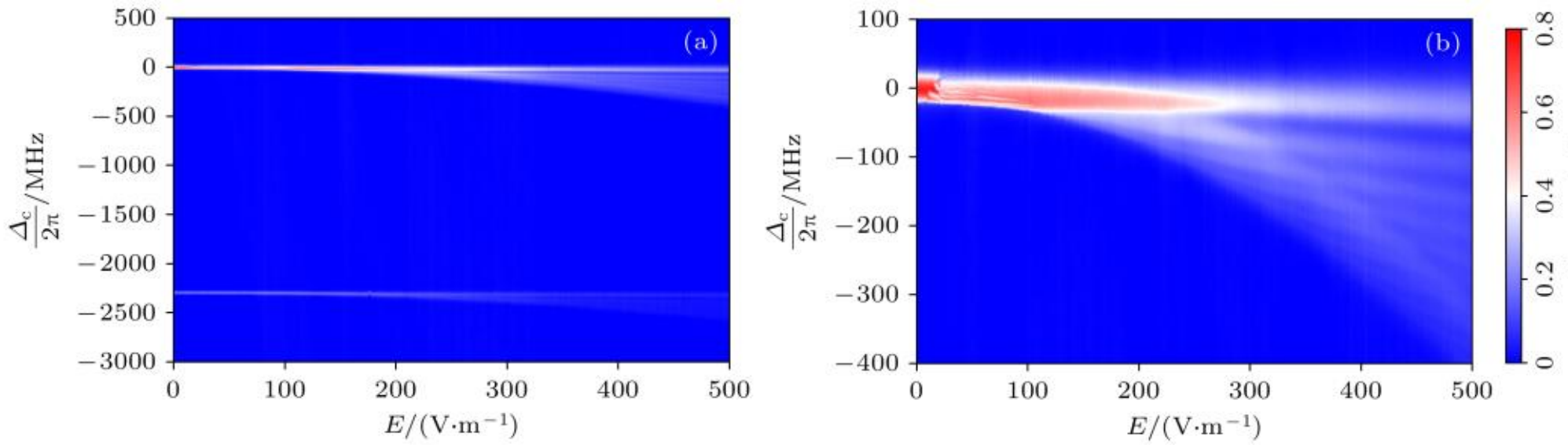

**Fig. 2 (a) EIT spectra of Rydberg atoms under 10 MHz sinusoidal radio frequency electric fields with varying intensities; (b) the EIT spectral characteristics of $|49P_{3/2}\rangle$ state Rydberg atoms in different intensities of radio frequency electric fields.**

The experiment further investigated the influence of radio-frequency (RF) electric fields on atomic EIT spectra. The RF electric field signal was applied to the electrode plates via a transmission line. When the signal source was activated, the frequency was set to 10 MHz, and the output power was 15 dBm. Upon activation of the RF electric field, the entire EIT spectrum shifted to the left by approximately 8 MHz, as shown in Fig. 3(b). The Stark level shift[4] satisfies $\delta$ = -(1/2)$\alpha E^2$, where $\alpha$ is the polarizability and $E$ is the electric field strength. Previous calculations have shown that the polarizability of Rydberg states is much larger than that of the ground and intermediate states, making Rydberg states more sensitive to external electric fields. When considering the perturbation of Rydberg states by an external electric field, the detuning of the coupling light in Eq. (1) can be corrected to $\Delta_c = \Delta_c + \delta$. When $\delta$ is negative, the EIT spectral line shifts significantly to the left. By utilizing the AC Stark effect, the electric field strength inside the cesium vapor cell can be calibrated. The polarizability of the Rydberg atom was calculated to be 1050.726 $MHz \cdot cm^2 \cdot V^{-2}$ using the ARC package[31]. When the Stark level shift is 8 MHz, the electric field strength is calculated to be 174 mV/cm. Due to the shielding effect of the vapor cell, the electric field strength experienced by the atoms is significantly lower than the voltage applied between the electrode plates. Meanwhile, we numerically calculated the transmission spectrum of the probe light for the $|49P_{3/2}\rangle$ state using the theoretical model described in Section 2.2. The parameters used were $\Gamma_{21} \rightarrow 2\pi\times5.234$ MHz, $\Gamma_{32} \rightarrow 2\pi\times3.288$ MHz, and $\Gamma_{43} \rightarrow 2\pi\times3.7$ kHz, based on the energy level parameters of cesium atoms. As shown in Fig. 3(a), the numerical results are approximately consistent with the experimentally measured EIT transmission spectra. Furthermore, the shift of the EIT spectral peak increases with the applied electric field strength. Additionally, the experimentally measured EIT spectra exhibit non-trivial phenomena such as phase transitions and oscillations. These effects are primarily caused by strong interactions between Rydberg atoms, which exceed the approximation conditions of the theoretical model. Methods such as the mean-field approximation need to be introduced to address these effects[34-37].

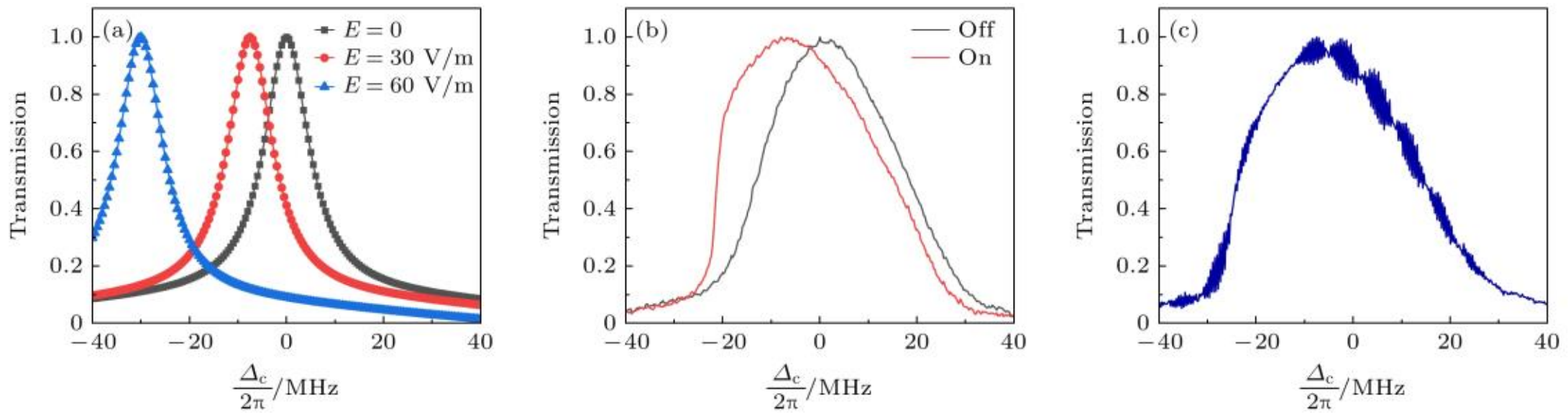

**Fig. 3 (a) Theoretically calculated EIT spectrum with applied electric field strengths of 0, 30, and 60 V/m; (b) activation of the radio frequency electric field at *P* = 15 dBm causes a noticeable leftward shift of approximately 8 MHz in the EIT spectrum. This change is clearly illustrated by comparing the black curve (deactivated state) with the red curve (activated state), demonstrating how the spectral position alters due to the applied RF field; (c) EIT spectrum observed when both the local oscillator field and the signal field are applied simultaneously. The spectral signal exhibits modulation due to the beat frequency of the two fields, with a frequency difference of $\Delta\omega$ = 100 kHz, where $f_{LO}$ = 10.1 MHz and $f_{sig}$ = 10 MHz; the transmittance depicted in the figure is normalized.**

To further enhance detection sensitivity, this study introduces a local oscillator electric field using the heterodyne method to measure weak megahertz radio-frequency electric fields based on Rydberg atoms. Low-frequency radio-frequency electric fields cause energy shifts in Rydberg states, which further shift the EIT spectrum of cesium atoms. When two electric fields of different frequencies are simultaneously applied to the electrode plates, they perturb the Rydberg energy levels to different extents, resulting in alternating features in the EIT spectrum. We activated two signal sources simultaneously, setting their frequencies to 10 MHz and 10.1 MHz, respectively. The resulting EIT spectrum is shown in Fig. 3(c). The EIT spectrum is clearly modulated and shifts to the left, with the measured modulation frequency equal to the difference between the frequencies of the local oscillator field and the signal field. The Stark level shifts induced by the local oscillator field and the signal field are

$$\delta = -\frac{1}{2}\alpha(E_{\mathrm{LO}} + E_{\mathrm{sig}})^2. \tag{4}$$

In the experiment, the local oscillator field and the signal field are set to have the same initial phase, namely $\varphi_{\mathrm{LO}} = \varphi_{\mathrm{sig}}$. After expanding Eq. (4), rapidly varying terms appear. By integrating to obtain the average value over one period, we obtain

$$\overline{\delta} = \overline{\delta}_0 - \frac{1}{2}\alpha[E_{\mathrm{LO}}E_{\mathrm{sig}}\cos(\Delta\omega \cdot t)], \tag{5}$$

Here, $\overline{\delta}_0 = -1/4\alpha(E_{\mathrm{LO}}^2 + E_{\mathrm{sig}}^2)$, represents the average Stark level shift induced by the local oscillator field and the signal field, while $\Delta\omega$ denotes the detuning between the local oscillator and the signal. The Stark level shift is modulated by the beat frequency in Eq. (5), allowing the beat signal to be observed in the EIT spectrum. We varied the coupling light detuning to optimize the beat signal intensity. Finally, we selected the detuning corresponding to the maximum beat response as the operating point for sensitivity detection. By measuring the beat frequency component of the probe light, we can extract information regarding the electric field strength in the signal. The outputs from two signal sources were applied to the electrode plates after passing through a radio-frequency power divider. The local oscillator frequency was set to 10.1 MHz with a power $P_{\mathrm{LO}}$ = 15 dBm, and the signal frequency was 10 MHz. The coupling light was fixed at the optimal operating point, and a digital oscilloscope monitored the temporal variation of the probe light intensity. Subsequently, a spectrum analyzer measured the beat signal intensity. Figure 4(a) shows the beat signal intensity under different signal electric field strengths. The received beat signal intensity is approximately proportional to the applied signal electric field strength, as described in Eq. (5). This setup achieved the measurement of a 10 MHz radio-frequency electric field. The dynamic range exceeded 65 dB, and the minimum detectable external input power reached -60 dBm. Using the calibration relationship between the electric field within the electrode plates and the input power, we calculated the electric field strength sensitivity to be 31.0 $\mu\mathrm{V\cdot cm^{-1}\cdot Hz^{-1/2}}$. Figure 4(b) illustrates the instantaneous bandwidth of the system. Considering the negative detuning between the signal field and the local oscillator field, the instantaneous bandwidth of this system reached 0.6 MHz. The system bandwidth primarily depends on the scattering rate of probe photons at the intermediate atomic resonance[14] (approximately 10 MHz). We define the parameter $\eta = B/\omega_c$, which is the ratio of bandwidth to carrier frequency, to characterize the relative bandwidth. Compared with experimental results in the resonant microwave band[3], Rydberg atoms demonstrate significantly improved bandwidth utilization for carrier waves in the low-frequency band. This finding highlights the advantage of Rydberg atoms for communication in low-frequency bands. This difference arises because previous studies employed electric fields in the GHz band to achieve resonant coupling of Rydberg levels. When a large deviation exists between the signal and local oscillator frequencies, the response capability of Rydberg atoms to the signal diminishes. In this work, we utilize the non-resonant AC Stark level shift effect, enabling atoms to exhibit a more significant bandwidth response to the electric field.

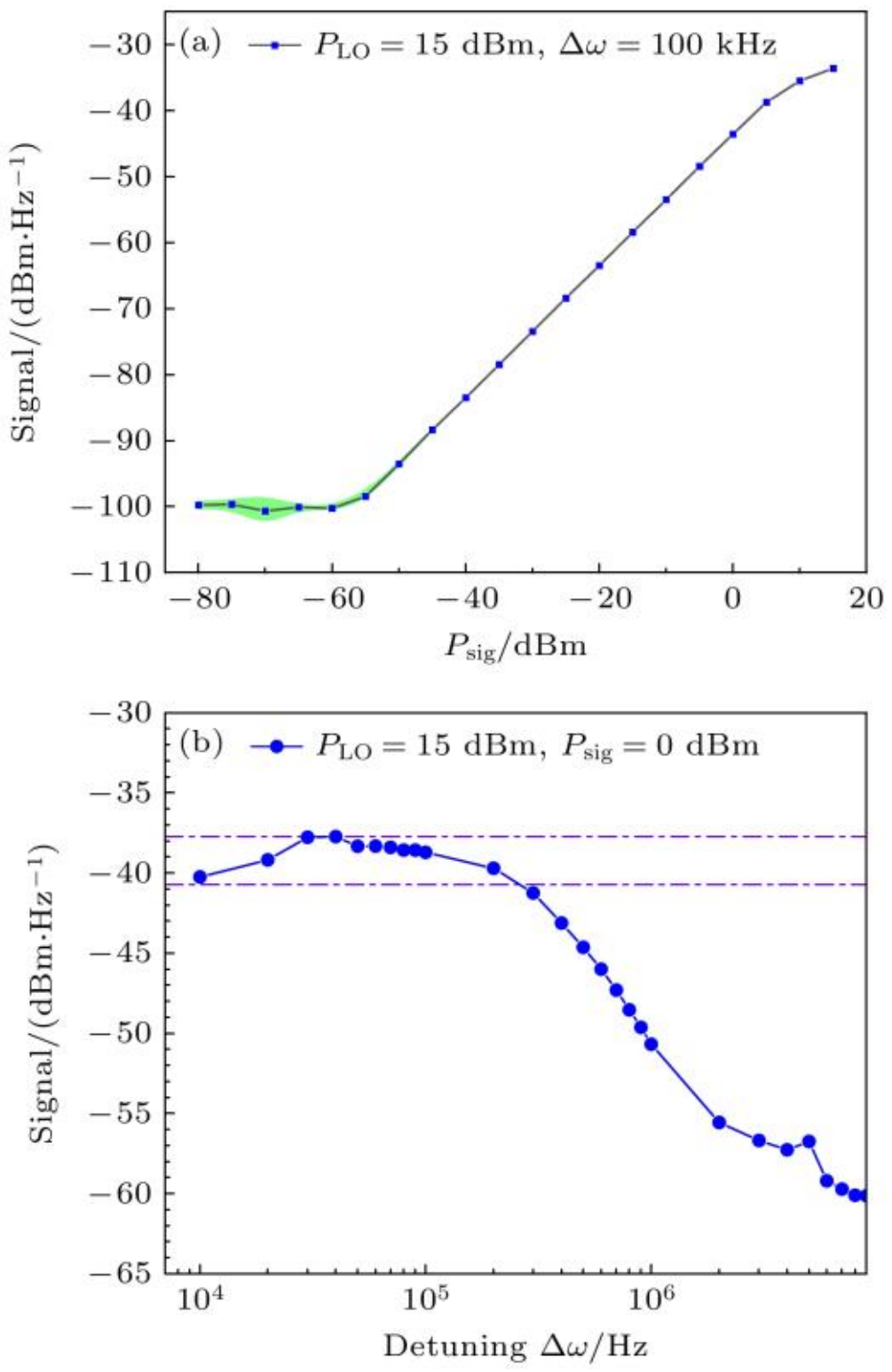

**Fig. 4 (a) The dynamic range of the system is measured at $P_{LO}$ = 15 dBm and $\Delta\omega$ = 100 kHz, where a linear relationship between the beat frequency signal strength and input power is observed over a 65 dB range; (b) the instantaneous bandwidth of the system was evaluated under operating conditions of $P_{LO}$ = 15 dBm and $P_{sig}$ = 0 dBm. As the detuning between the signal electric field and the local oscillator electric field increases, the intensity of the beat frequency signal decreases continuously. Considering the negative detuning scenario, a 3-dB bandwidth of 0.6 MHz was measured.**

# 4 Conclusion

We constructed a ladder-type four-level system of cesium atoms ($|6S_{1/2}\rangle \rightarrow |6P_{3/2}\rangle \rightarrow |7S_{1/2}\rangle \rightarrow |49P_{3/2}\rangle$) at room temperature. This setup enabled continuous observation and recording of Rydberg atom EIT spectra under the dressing of a 10 MHz radio-frequency electric field. By analyzing the perturbation mechanism of megahertz-band electric fields on Rydberg energy levels and combining this with heterodyne detection technology, we achieved high-sensitivity measurement of weak radio-frequency electric fields. Experimental results indicate that the system achieves an electric field sensitivity of 31.0 $\mu V \cdot cm^{-1} \cdot Hz^{-1/2}$ at 10 MHz, with a dynamic range of 65 dB. Based on the quantum sensing characteristics of Rydberg atomic systems, this technology demonstrates significant application potential in the shortwave communication band (3-30 MHz). Its centimeter-scale miniaturized structure and high-sensitivity response capability are poised to overcome the trade-off between size and sensitivity inherent in traditional antennas. Consequently, it provides a new technical pathway for

miniaturized electric field sensors in long-range wireless communication, maritime navigation, and aviation communication systems.